\documentclass[a4paper,fleqn]{cas-sc}

\usepackage[numbers]{natbib}
\usepackage{graphicx}
\usepackage{subcaption}
\usepackage{float}
\usepackage{siunitx}

\def\tsc#1{\csdef{#1}{\textsc{\lowercase{#1}}\xspace}}
\tsc{WGM}
\tsc{QE}
\tsc{EP}
\tsc{PMS}
\tsc{BEC}
\tsc{DE}

\begin{document}
\let\WriteBookmarks\relax
\def\floatpagepagefraction{1}
\def\textpagefraction{.001}
\shorttitle{Overview of the ALICE ITS3 Upgrade}
\shortauthors{N. Bouchhar}

\title [mode = title]{Overview of the ALICE ITS3 Upgrade}                      

\author[1]{Naseem Bouchhar}[orcid=0000-0002-5129-5705]
\affiliation[]{On Behalf of the ALICE Collaboration}
\affiliation[1]{organization={Department of Physics and Astronomy, Sejong University},
                city={Seoul},
                country={South Korea}}

\begin{abstract}
\noindent The ALICE experiment will replace its three innermost tracking layers with the Inner Tracking System 3 (ITS3) during LHC Long Shutdown 3. This upgrade introduces the first fully cylindrical, wafer-scale silicon vertex detector, utilising Monolithic Active Pixel Sensors (MAPS) fabricated in a \qty{65}{\nano\metre} CMOS process. By thinning sensors to \qty{50}{\micro\metre} and bending them to radii as small as \qty{19}{\milli\metre}, the design achieves a self-supporting structure that eliminates traditional support material. Wafer-scale stitching enables \qty{27}{\centi\metre}-long seamless sensors with integrated power and signal distribution, removing the need for flexible printed circuits within the active volume. These innovations, combined with a move from water to air cooling, reduce the material budget to less than 0.09\%X$_0$ per layer.

The R$\&$D program has been validated through full-scale prototypes (MOSS, MOST), which demonstrated stitching feasibility, high yield, and radiation hardness. Engineering models confirmed the feasibility of air-convection cooling, indicating effective thermal management and structural stability. This contribution summarises the key advances in stitched sensor development, mechanical integration, and the path toward the final qualification model.
\end{abstract}


\maketitle

\section{Introduction} 
\label{sec:Introduction}

\noindent To prepare for the Run 3 data-taking period, the ALICE experiment underwent a major upgrade during Long Shutdown 2 (LS2). Integral to this was the installation of the ITS2, a silicon-based vertex tracker featuring a three-layer inner barrel and a four-layer outer barrel~\cite{ITS2_TDR}. These layers are built from stave support structures that house nine sensors each in the inner barrel. To function, these staves require several layers of infrastructure including flexible printed circuit boards (FPCBs) for data and control transmission, and water-cooled plates to manage heat dissipation. In total, the ITS2 inner barrel contains 432 sensors, each \qty{50}{\micro\metre} thick.

\noindent Looking towards the third long shutdown (2026–2030), ALICE plans to replace the three innermost layers of ITS2 with a next-generation upgrade: ITS3~\cite{ITS3_TDR}. The primary goal of this upgrade is to drastically reduce the material budget. Building on the already excellent performance of ITS2, minimising this material is crucial in further elevating the detector's pointing resolution, which is the detector's ability to accurately track short-lived particles like charm and bottom quarks, particularly at low momenta.

\medskip

\noindent ITS3 achieves this reduction through several key innovations:
\begin{itemize}
    \item \textbf{Structural Geometry}: By using wafer-scale bent sensors, the structure becomes self-supporting. This allows for the removal of the stave supports, leaving only minimal carbon foam along the edges of the design.
    
    \item \textbf{Circuit Integration}: By moving from a \qty{180}{\nano\metre} to a \qty{65}{\nano\metre} fabrication process, the power and data electronics are integrated directly onto the silicon chips, eliminating the need for bulky power buses and FPCBs.

    \item \textbf{Air Cooling}: Shifting from water to air cooling removes the need for cooling plates and eliminates the material mass of the water itself.
\end{itemize}

\section{ITS3 Upgrade} 
\label{sec:ITS3}

\begin{figure}[pos=!htp]
    \centering

    \begin{subfigure}{0.54\linewidth}
        \centering
        \includegraphics[width=\linewidth]{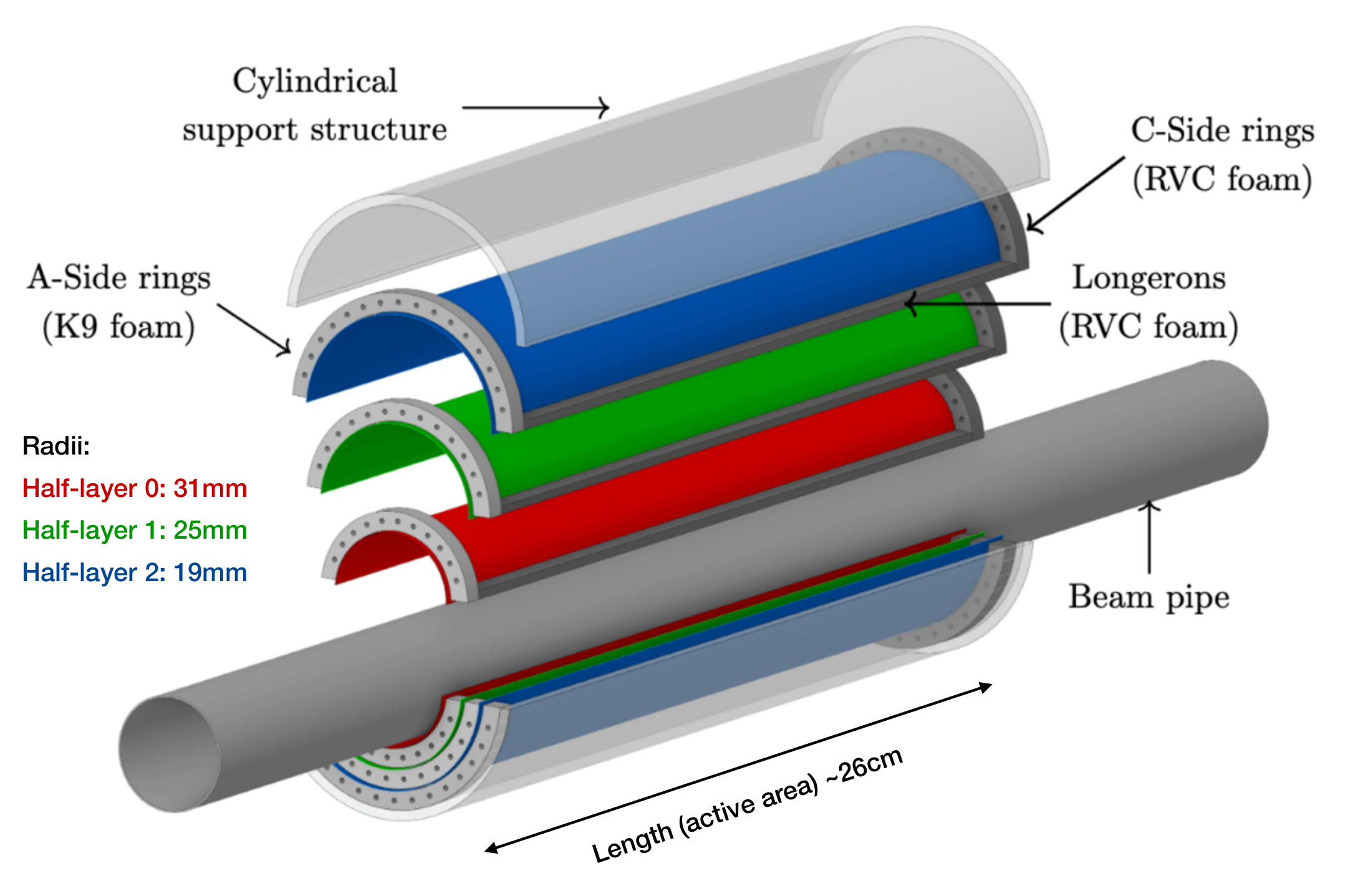}
        \caption{}
        \label{fig:ITS3_a}
    \end{subfigure}
    \hfill
    \begin{subfigure}{0.44\linewidth}
        \centering
        \includegraphics[width=\linewidth]{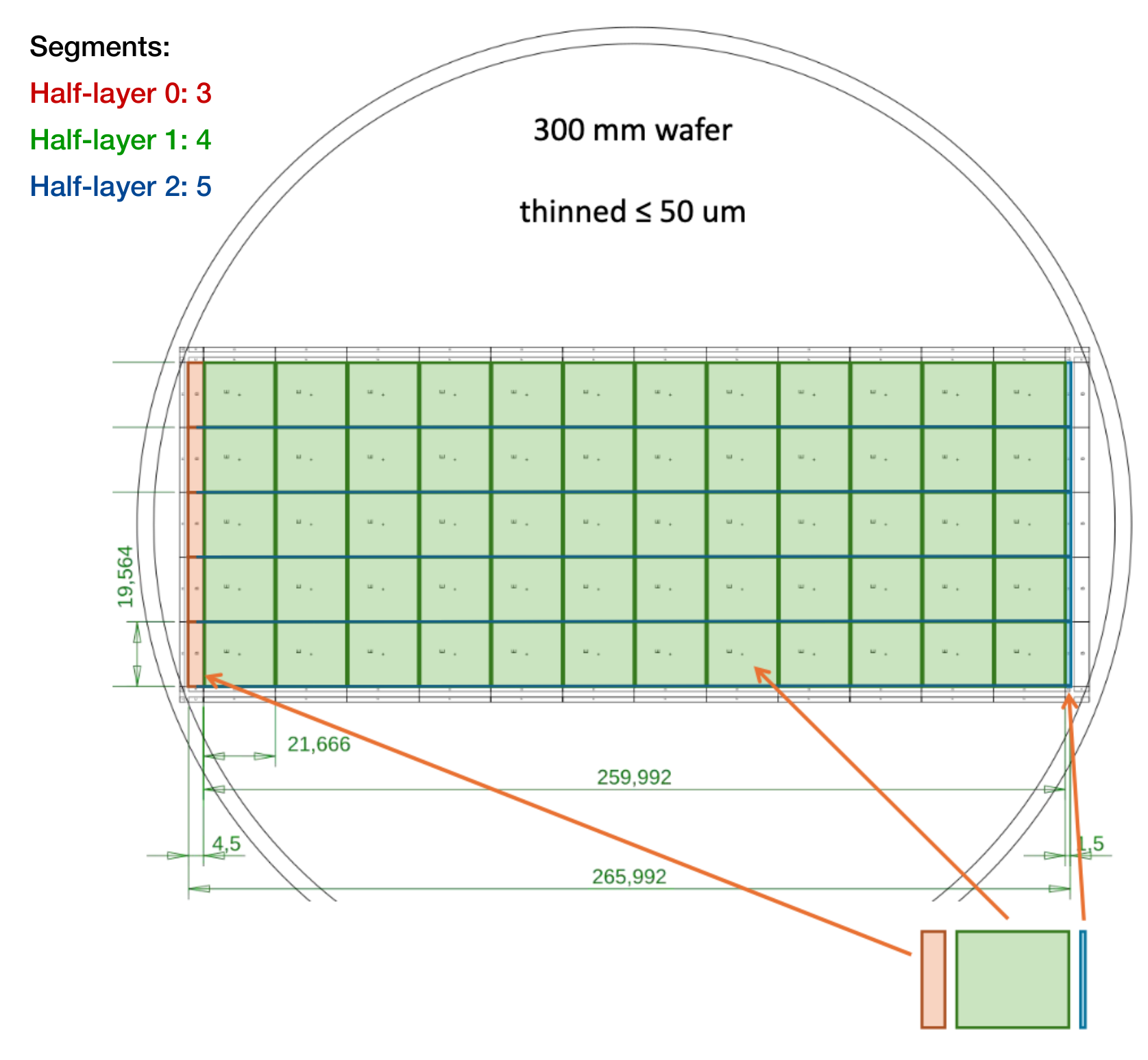}
        \caption{}
        \label{fig:ITS3_b}
    \end{subfigure}

    \caption{(a) ITS3 design, and (b) the corresponding wafer for the ITS3 sensors.}
    \label{fig:ITS3_design}
\end{figure}

\noindent The ITS3 upgrade for the ALICE experiment represents a fundamental shift in vertex detector technology, moving from traditional modular silicon assemblies to `bent-wafer' architecture. Figure~\ref{fig:ITS3_a} shows the planned ITS3 design, featuring six half-layers of bent, stitched MAPS sensors arranged around the beampipe.

\newpage

\noindent \textbf{Structural Geometry}

\smallskip

\noindent The mechanical realisation of the ITS3 relies on the inherent flexibility of silicon when thinned to \qty{50}{\micro\metre}, allowing the sensors to be physically bent into semi-cylindrical shapes around the beam pipe. This allows for a self-supporting structure, where the closest layer will be \qty{19}{\milli\metre} from the interaction point, a decrease of \qty{4}{\milli\metre} compared to ITS2. Because each layer is positioned at a different radial distance from the interaction point, their widths are adjusted to ensure complete coverage. Figure~\ref{fig:ITS3_b} illustrates the wafer-scale layout of the sensors comprising ITS3, where each half-layer requires a different number of wafer segments to accommodate these varying widths. Despite this variation, the longitudinal length along the beam direction ($z$) is kept constant at \qty{266}{\milli\metre} for all layers, making efficient use of nearly the full diameter of the production wafer.

\medskip

\noindent \textbf{Circuit Integration}

\smallskip

\noindent The detector requires wafer-scale sensor ASICs where each half-layer is constructed as a single, continuous piece of silicon. The transition from the \qty{180}{\nano\metre} CMOS technology used in the ITS2 to a \qty{65}{\nano\metre} CMOS process is the primary enabler of this design. Beyond simply allowing for higher transistor density and more complex in-pixel logic, the \qty{65}{\nano\metre} process is typically fabricated on \qty{300}{\milli\metre} wafers. This larger substrate is essential for manufacturing sensors that reach the \qty{266}{\milli\metre} length required for the ITS3 geometry, a feat that would not be possible on the smaller \qty{200}{\milli\metre} wafers used for older technologies.
However, because the dimensions of these half-layer sensors exceed the standard reticle size of lithographic equipment, a technique known as stitching is used. This process allows multiple design units to be electrically connected during fabrication, enabling the routing of power distribution and signals directly on the sensor chips through a stitched backbone. Consequently, the need for external FPCs is eliminated, removing a significant source of non-active material from the detector volume.

\medskip

\noindent \textbf{Mechanics and Air Cooling}
\smallskip

\noindent While ITS2 relies on water-cooled plates to dissipate sensor heat, ITS3 will make use of cooling using air convection.

To manage temperature gradients in the endcap regions, the design incorporates two types of carbon foam. The first, Carbon (RVC) Duocel$^\copyright$ manufactured by ERG Aerospace, is used to form open-cell foam rings placed in direct contact with the silicon sensors. These rings act as heat exchangers, conducting heat away from the sensors and dissipating it through air convection. This foam was selected for its high Young's modulus, low density, and isotropic thermal properties, which together ensure both structural integrity and uniform heat distribution. The second, Allcomp K9 standard density foam developed by Lockheed Martin, forms supports along the long edges of the detector to maintain the sensors' cylindrical geometry. This specific foam is dense with tiny cells and infused with graphite, creating a high surface area that allows it to absorb and transfer heat efficiently.

To validate this approach, a breadboard model was constructed using dummy silicon sensors equipped with polyamide heaters and copper patterns to simulate real-world heat production. This model was tested in a wind tunnel to determine the airflow required to stabilise the sensors. Given the current power estimate of approximately 40 mW/cm$^2$, the study found that a freestream air velocity greater than 5 m/s is sufficient to meet the design requirement, maintaining a maximum temperature variation of less than 10 K across the pixel matrix~\cite{Amatriain:2023bja}.

Additionally, the impact of airflow-induced vibrations was evaluated, as shifting pixel positions due to vibrations could have a degrading effect on spatial resolution. Using confocal chromatic sensors to measure physical displacement, it was found that, with an airspeed of 8 m/s, the maximum displacement was less than \qty{0.5}{\micro\metre}~\cite{Amatriain:2024ols}. This comfortably meets the ITS3 design requirement, which permits a maximum vibration of \qty{2}{\micro\metre}.

\medskip

\begin{figure}[pos=!htb]
    \centering

    \begin{subfigure}{0.55\linewidth}
        \centering
        \includegraphics[width=\linewidth]{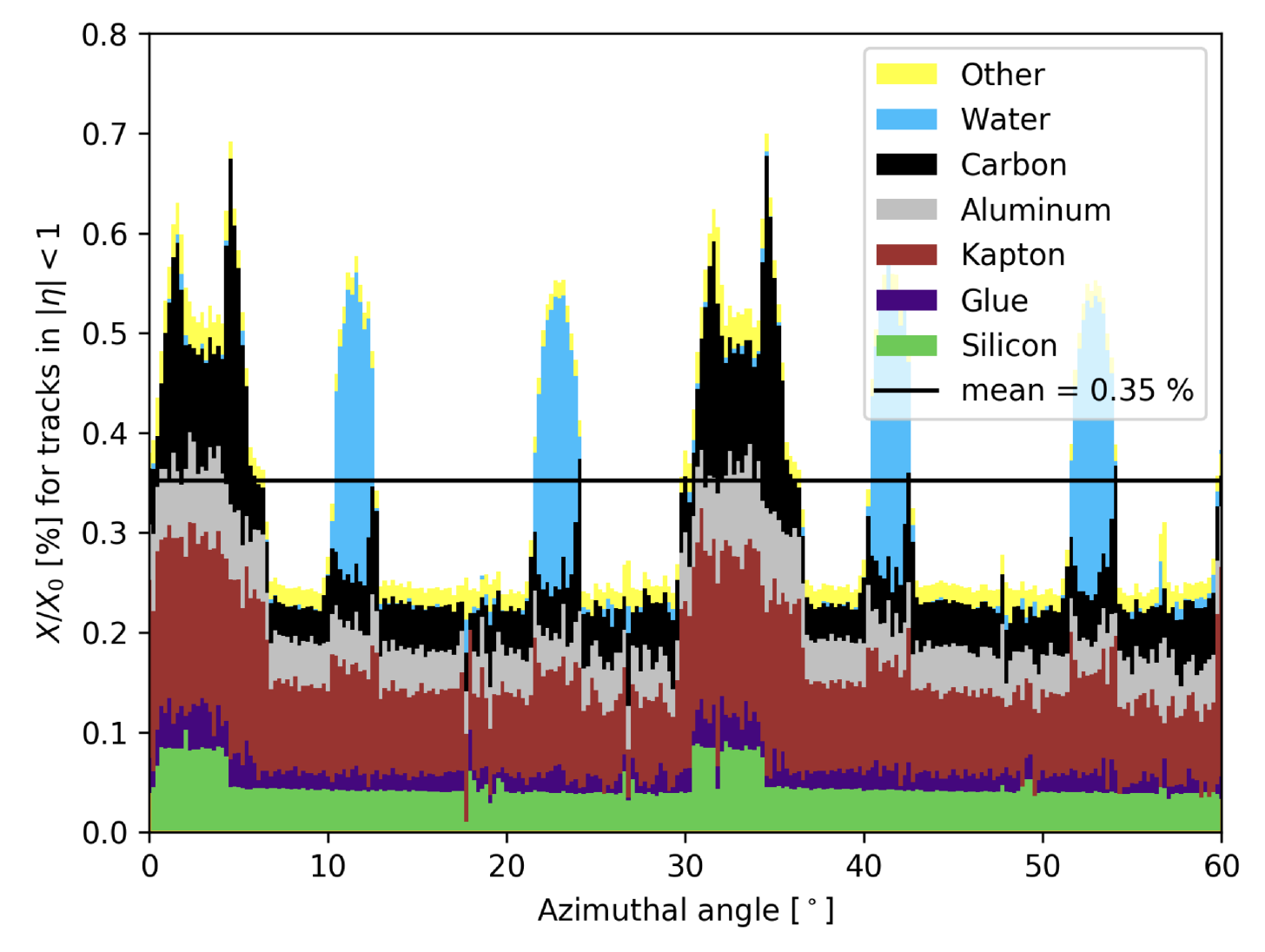}
        \caption{}
        \label{fig:mat_budg_ITS2}
    \end{subfigure}
    \hfill
    \begin{subfigure}{0.44\linewidth}
        \centering
        \includegraphics[width=\linewidth]{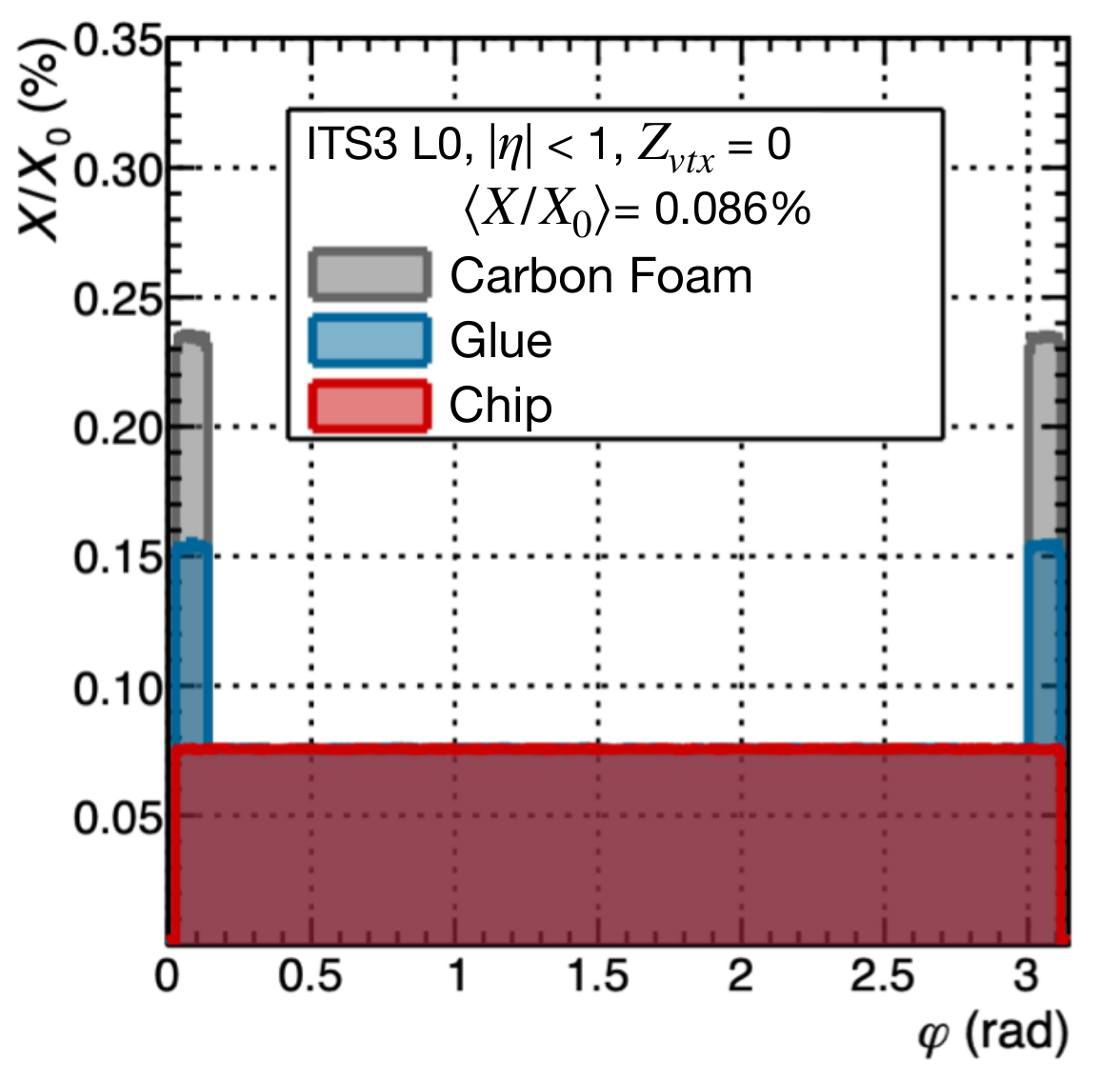}
        \caption{}
        \label{fig:mat_budg_ITS3}
    \end{subfigure}

    \caption{Material budget per layer as a function of azimuthal angle, $\phi$, for (a) ITS2 and (b) ITS3.}
    \label{fig:material_budget}
\end{figure}

\noindent The overall result of these improvements is a system of self-supporting silicon sensors that are virtually free of non-active material. This design reduces the material budget from 0.36\% down to around 0.09\% per layer, as shown in Figure~\ref{fig:material_budget}.  By integrating all electrical services onto the silicon itself and utilising air-cooling, the material distribution remains almost perfectly uniform. The only additional elements within the active volume are minimal amounts of glue and ultra-lightweight carbon fibre support structures, ensuring that particle trajectories are measured with minimal interference.

\section{Prototype Sensors: MOSS and MOSAIX} 
\label{sec:MOSS}

\begin{figure}[pos=!htb]
    \centering
    \includegraphics[width=\linewidth]{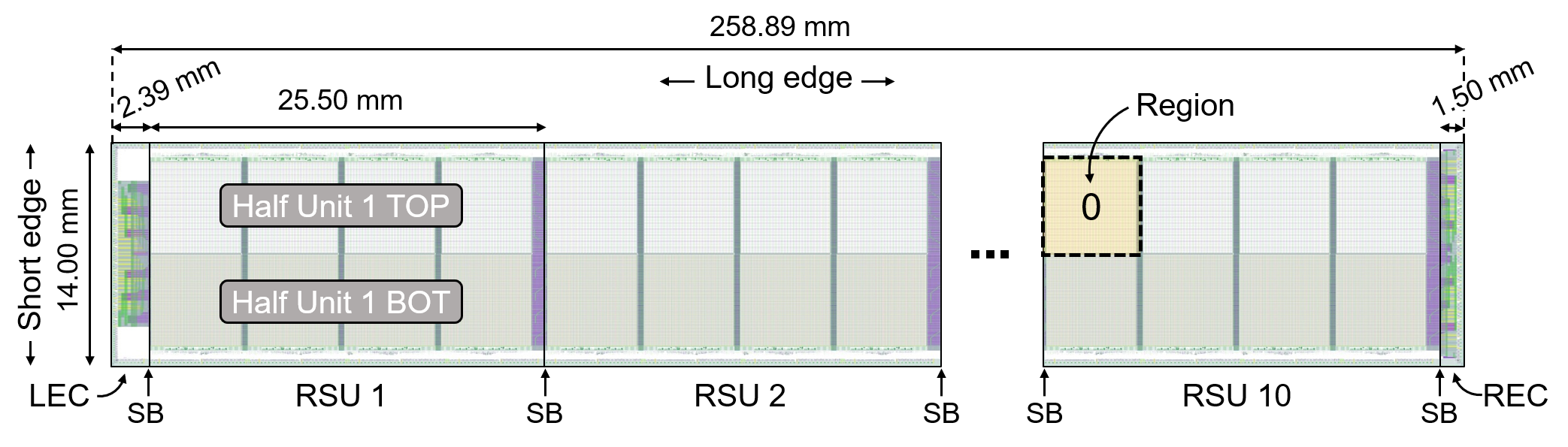}
    \caption{Architecture of the MOSS prototype sensor.}
    \label{fig:moss_architecture}
\end{figure}

\noindent The first full-scale validation of the MAPS stitching technology was realised through Engineering Run 1 (ER1) with the MOnolithic Stitched Sensor (MOSS), the architecture of which is illustrated in Fig. \ref{fig:moss_architecture}. This prototype primarily serves to demonstrate the viability of the stitching process and ensure it meets the stringent requirements for the ITS3 upgrade.

The MOSS sensor comprises 10 Repeated Sensor Units (RSUs) that, while functionally identical, are designed for independent power and readout. Each RSU is subdivided into eight regions, four top and four bottom, featuring various pixel matrix designs and pitches of \qty{22.5}{\micro\metre} and \qty{18}{\micro\metre}, respectively. This variety facilitates the optimisation of transistor sizes and pixel geometry for further ITS3 sensor production. MOSS supports two distinct operational modes: the first provides independent control of each half-unit via bond pads along the sensor's long edge, and the second utilises Input/Output (I/O) blocks within the Left Endcap (LEC).

To mitigate the conductivity limits of the metal stack, power is supplied directly through bond pads at the top and bottom of each individual RSU. This distributed supply approach prevents significant voltage drops across the \qty{27}{\centi\metre} sensor length.

A comprehensive characterisation campaign was conducted on 82 MOSS sensors~\cite{Abdelrahman:2025vkk}, yielding the following key results:
\begin{itemize}
    \item \textbf{Yield and Reliability:} Excluding metal stack defects caused by manufacturing imperfections, which affected approximately 14\% of RSUs, the laboratory yield for MOSS was about 98\%. These metal stack issues were reported to the foundry to help reduce such defects in future production runs. Overall, the high yield indicates a robust design with consistent performance across wafers and no significant difference in failure rates between top and bottom RSU configurations.
    \item \textbf{Beam Test Performance:} In-beam measurements confirmed a detection efficiency above 99\% and a fake-hit rate below 10$^{-1}$ hits/pixel/s. Measured spatial resolutions (4-\qty{5.5}{\micro\metre}) across the different pixel pitches validate that an intermediate pixel size will meet the ITS3 target of \qty{5}{\micro\metre}, with no significant performance variation between the pixel design variants.
\end{itemize}

\medskip

\noindent ER2 introduces the MOnolithic Stitched Active pIXel (MOSAIX) sensor, which serves as the functional prototype for the final design of the ITS3 detector sensors.

As opposed to the MOSS, power is distributed through only the LEC and REC, while the LEC alone manages control signals and data readout via eight high-speed channels supporting data rates up to 10.24 Gb/s. MOSAIX sensors are divided into 12 RSUs, subdivided into 12 independent tiles per RSU, arranged in two rows of 6. Each tile is a self-contained sensor block with its own power domain and control interface, allowing for selective power switching and precise management of the detector.

The pixel matrix within each tile consists of 444 rows and 156 columns with a pitch of 20.8 by \qty{22.8}{\micro\metre}. By implementing 12 distinct pixel geometries across the RSUs, the MOSAIX will be used to finalise an optimal in-pixel design for the final production-grade sensors.

\section{Summary} 
\label{sec:Summary}

\noindent The ALICE ITS3 upgrade will replace the inner barrel of ITS2 with ultra-thin, self-supporting, wafer-scale sensors thinned to only 50 micrometers. This removes the need for external circuit boards and heavy support frames, and the low power consumption of the sensors allows for air cooling instead of water. These innovations will reduce the material budget significantly, thus improving the vertexing capabilties of the ALICE experiment.
The ER1 prototype successfully validated the wafer-scale stitching process, with the sensors providing a laboratory yield of approximately 98\%. While MOSS confirmed the feasibility of signal and power routing across large silicon areas, the ER2 prototype sensor is designed to provide the final technical qualification. MOSAIX will validate the integrated data and power distribution, with high-speed data rates, while finalising the optimal pixel design for the production-grade detector. 

\printcredits

\bibliographystyle{cas-model2-names}

\bibliography{cas-refs}

\end{document}